\documentclass[aip,pop,preprint,a4paper]{revtex4-1}

\usepackage{graphicx}

\begin{document}

\title{Gyrokinetic Simulations on Zonal Flow--Turbulence Spreading Coupling}

\author{M.K. Jung}
\affiliation{Department of Nuclear Engineering, Seoul National University, Seoul 08826, South Korea}
\author{S. Yi}
\affiliation{Korea Institute of Fusion Energy, Daejeon 34113, South Korea}
\author{T.S. Hahm}
\email[]{tshahm@snu.ac.kr}
\affiliation{Nuclear Research Institute for Future Technology and Policy, Seoul National University, Seoul 08826, South Korea}
\author{Y.-S. Na}
\affiliation{Department of Nuclear Engineering, Seoul National University, Seoul 08826, South Korea}


\begin{abstract}
Zonal flows and turbulence spreading play important roles in magnetic fusion plasma confinement, yet their coupling mechanisms remain elusive. Using global nonlinear gyrokinetic simulations, we show that turbulence spreading transports zonal flow radially, extending into the linearly stable regions after local nonlinear saturation of turbulence. Theoretical understanding has been gained by analyzing the simulation results in the context of a momentum theorem in toroidal plasmas [T.S. Hahm \textit{et al.}, Phys. Plasmas \textbf{31}, 032310 (2024)] which is an extension of the Charney-Drazin non-acceleration theorem [J.G. Charney and P.G. Drazin, J. Geophys. Res. \textbf{66}, 83 (1961)]. It indicates a direct relation between turbulence-driven enstrophy transport and perpendicular momentum generation.
\end{abstract}

\maketitle

\section{Introduction}\label{sec:introduction}

Importance of both zonal flows (ZF) and turbulence spreading (TS) in magnetic fusion plasma confinement is widely recognized as summarized in review articles \cite{DiamondPPCF05,HahmJKPS18} respectively. While most global gyrokinetic simulations have observed both turbulence-driven zonal flows \cite{LinSci98} and turbulence spreading in the past, \cite{GarbetNF10} their physics has been discussed separately in most cases.

Studies on turbulence spreading which were initiated in Ref.~\onlinecite{GarbetNF94} have been applied to the size scaling of confinement issue (i.e., deviation from the gyroBohm scaling) \cite{LinPRL02,HahmPPCF04} and the turbulence short fall problem in the edge-core connection zone (so called the no-man's land). \cite{HahmPoP05,CaoPRL25} In addition, an extended version of Fisher-Kolmogorov equation \cite{FisherAE37,KolmogoroffCCR37} has been adopted as theoretical model \cite{HahmPPCF04,GurcanPoP05} without including the effect of zonal flows. Shortly after those studies on turbulence spreading, role of strong equilibrium $E\times B$ shear layer in blocking turbulence spreading has been demonstrated using global gyrokinetic simulations. \cite{WangPoP07} There are experimental evidences from stellarators of the radial electric field transition at the edge blocking the radial propagation of turbulence from edge to Scrape-Off-Layer \cite{GrenfellNF20}, and an increase of turbulence in the inner region as the flow shear decreases at the boundary. \cite{EstradaNF11} In addition, low-mode number magnetic fluctuations can destroy the $E\times B$ shear layer at H-mode plasma edge, and it can lead to a beneficial effect of reducing a peak value of divertor heat load. \cite{KobayashiPRL22} On the other hand, zonal flow generation has been usually studied in the context of modulational instability \cite{ChenPoP00,GuzdarPoP01,ShuklaEPJD02,HahmPoP23} without considering the effect of turbulence spreading.

The first theoretical attempt which explicitly considered the effect of turbulence-driven zonal flows (ZF) on turbulence spreading was Ref.~\onlinecite{ChenPRL04} to our knowledge. Their results indicate that the zonal flows promote turbulence spreading. This conclusion sounds counter-intuitive because one expects zonal flows without radial component should not cause transport of either particles or fluctuation energy of drift waves (DW). This can be understood by observing that Ref.~\onlinecite{ChenPRL04} considers only the nonlinear mode coupling channels involving DW-ZF-DW. Later, Refs.~\onlinecite{GurcanPRL06} and \onlinecite{GurcanPoP06} stated that based on a certain nonlinear model \cite{ChenPRL04} one may get misled to believe that zonal flow is the cause of turbulence spreading. In such models, if one turns off the zonal flow, there will be no nonlinear interactions and therefore no turbulence spreading. Obviously, the relation between ZF and TS is a highly relevant but subtle issue.

In this work, we pursue understanding the relation between zonal flow evolution and turbulence spreading from global nonlinear gyrokinetic simulations using \texttt{gKPSP} \cite{KwonNF12,KwonCPC17} code. Our gyrokinetic simulations presented in this paper demonstrate that after a local nonlinear saturation of turbulence, turbulence spreading radially carries zonal flows alongside. This trend has been observed in a previous \texttt{gKPSP} simulation study, \cite{YiPoP24} although comparisons to theory have not been performed. In this paper, simulation results are analyzed in the context of the prediction from the momentum theorem for toroidal plasmas \cite{HahmPoP24} which can be derived from the modern nonlinear gyrokinetic equation \cite{HahmPF88} in the cold ion limit. Such conservation laws, rather than particular nonlinear mode coupling calculations, can provide more definite conclusions on this subtle issue. This is an extension of the Charney-Drazin (CD) non-acceleration theorem \cite{CharneyJGR61} which is well-known in geophysical fluid dynamics (GFD) community. \cite{VallisBook} The first extension of the CD theorem for nuclear fusion plasma application \cite{DiamondPPCF08} considered the Hasegawa-Wakatani equation. \cite{HasegawaPRL83}

The remainder of this paper is organized as follows. Sec.~\ref{sec:background}. summarizes the physics of momentum theorem in toroidal plasmas which will be used for understanding the gyrokinetic simulation results. Details of \texttt{gKPSP} global nonlinear gyrokinetic simulations are presented in Sec.~\ref{sec:simulation}. Further discussions are given and conclusions are drawn in Sec.~\ref{sec:conclusion}.

\section{Theoretical Background}\label{sec:background}

In this section, we briefly sketch the derivation of the momentum theorem in toroidal plasmas \cite{HahmPoP24} and present the key expression which will be used for validations against the \texttt{gKPSP} simulation results. The momentum theorem can be derived for a system conserving a potential vorticity (PV). PV dynamics is closely related to the zonal flow evolution, and inhomogeneous mixing of PV is a candidate mechanism for zonal flow shear layers. \cite{McIntyreJMSJ82}

For our purpose, it is useful to recall that PV conservation in magnetized plasma can be considered by starting from the ion gyrocenter density continuity equation,
\begin{equation}\label{eqn1}
    \frac{\partial}{\partial t}n_G+\vec{\nabla}\cdot(n_G\vec{u}_E)=0,
\end{equation}
where $\vec{u}_E=\frac{c\hat{b}\times\vec{\nabla}\delta\phi}{B}$. This form is strictly valid in 2d cold-ion limit in which $u_\parallel$ and $u_{di}$ (grad-$B$ and curvature drifts) do not appear. From gyrokinetic Poisson equation (quasi-neutrality condition for gyrokinetic system), $n_G=n_e-n_{pol}$, where $n_{pol}$ is the ion polarization density arising from the difference between the particle density and the gyrocenter density. \cite{LeePF83} When one takes inhomogeneity of density and magnetic field into account, the polarization density is given by \cite{StrintziPoP04,HahmPoP09,MadsenPoP13,WangPoP09}
\begin{equation}\label{eqn2}
    n_{pol}=\nabla_\perp\cdot\frac{n_0M_ic^2}{|e|B^2}\nabla_\perp\delta\phi.
\end{equation}
Therefore, in terms of dimensionless variables, the potential vorticity $q$ is given by
\begin{equation}\label{eqn3}
    q=N_e-\nabla_\perp\cdot\left(\frac{N}{B^2}\nabla_\perp\Phi\right),
\end{equation}
where $N_e=n_e(\vec{x})/n_0$, $N=n_0(\vec{x})/n_0$, $B=B_{phys}(\vec{x})/B_0$, $\Phi=|e|\delta\phi(\vec{x})/T_e$, and $\nabla_\perp$ is normalized to the inverse of $\rho_s\equiv c_s/\Omega_{ci}=(T_e/M_i)^{1/2}/(|e|B_0/M_ic)$.

Eq.~(\ref{eqn1}) leads to
\begin{equation}\label{eqn4}
    \frac{\partial}{\partial t}q+\vec{\nabla}\cdot(q\vec{u}_E)=0.
\end{equation}
For a low-$\beta$ magnetic equilibrium, Eq.~(\ref{eqn1}) can be written as \cite{McDevittPoP10}
\begin{equation}\label{eqn5}
    \left(\frac{\partial}{\partial t}+\vec{u}_E\cdot\vec{\nabla}\right)q_M=0,
\end{equation}
where the magnetic potential vorticity (MPV) is given by
\begin{equation}\label{eqn6}
    q_M\equiv\frac{\left(N_e-\nabla_\perp\cdot\frac{N}{B^2}\nabla_\perp\Phi\right)}{B^2}.
\end{equation}
Eq.~(\ref{eqn5}) states that MPV is conserved along the path given by the turbulent field $\vec{u}_E$. As a consequence, the inhomogeneous mixing of MPV can lead to a formation of zonal flow shear layers. \cite{DritschelJAS08}
Since we are considering ITG turbulence only in this paper, an adiabatic response is assumed for electrons. So we have $\delta n_e=|e|\frac{\delta\phi-\langle\delta\phi\rangle}{T_e}n_0$, or $N_e=N_0\left(1+(\phi-\langle\phi\rangle)\right)$ in dimensionless form. Here, $\langle \delta\phi\rangle\equiv\oint{dy\delta\phi}/\oint dy$ is flux-surface average where $y$ is the standard bi-normal coordinate.

A key step in the derivation of momentum theorem is the following. By flux-surface-averaging the perturbed part of Eq.~(\ref{eqn4}), one can obtain
\begin{equation}\label{eqn7}
    \frac{\partial}{\partial t}\left\langle\nabla_\perp\cdot\frac{N}{B^2}\vec{\nabla}_\perp\Phi\right\rangle=\Bigl\langle\vec{\nabla}\cdot(\delta q\vec{u}_E)\Bigr\rangle.
\end{equation}
The PV flux is proportional to the Reynolds stress in slab geometry or $\beta$-plane model in GFD, known as the Taylor's identity, \cite{TaylorPTRSA15}
\[
    \langle\nabla_\perp^2\phi u_r\rangle=\frac{\partial}{\partial r}\langle u_\theta u_r\rangle.
\]
Although its extension to toroidal geometry has not been achieved to our knowledge, the simple relation above has been used in interpreting simulation and experimental results in toroidal geometry as well in the past.

Rather than using the Taylor's identity, we follow a procedure outlined in Ref.~\onlinecite{HahmPoP24}. By observing the fact that only the radial derivatives of $\frac{N}{B^2}\nabla_\perp\Phi$ and $\delta q\vec{u}_E$ survive the flux-surface-average in Eq.~(\ref{eqn7}), we integrate once in $\psi$ (flux function designating the radius) to get
\begin{equation}\label{eqn8}
    \frac{\partial}{\partial t}\left\langle\frac{N}{B^2}\vec{\nabla}\Phi\cdot\vec{\nabla}\psi\right\rangle=\langle\delta q\vec{u}_E\cdot\vec{\nabla}\psi\rangle.
\end{equation}
Meanwhile, one can also construct an evolution equation for $\langle\delta q^2\rangle$ following Ref.~\onlinecite{HahmPoP24}.
\begin{equation}\label{eqn9}
    \frac{1}{2B^2}\frac{\partial}{\partial t}\langle\delta q^2\rangle+\langle\delta q\vec{u}_E\cdot\vec{\nabla}\psi\rangle\frac{\partial}{\partial \psi}\left(\frac{N}{B^2}\right)+\frac{1}{2V'}\frac{\partial}{\partial \psi}\Bigl(V'B^2\langle\delta q_M^2\vec{u}_E\cdot\vec{\nabla}\psi\rangle\Bigr)=0.
\end{equation}
Equilibrium magnetic field is given as
\begin{equation}\label{eqn10}
    \vec{B}=\vec\nabla\zeta\times\vec\nabla\psi+I(\psi)\vec\nabla\zeta=B_\theta\hat e_\theta+B_\zeta\hat e_\zeta,
\end{equation}
where $\vec\nabla\psi=RB_\theta\hat e_\psi$ and $I(\psi)=RB_\zeta$. Here, flux coordinates $(\psi,\theta,\zeta)$ are used for the radial, poloidal, and toroidal coordinates. Then, in this coordinate system,
\[
    \vec\nabla=\hat e_\psi RB_\theta\frac{\partial}{\partial\psi}+\hat e_\theta\frac{1}{JB_\theta}\frac{\partial}{\partial\theta}+\hat e_\zeta\frac{1}{R}\frac{\partial}{\partial\zeta},
\]
where the Jacobian $J\equiv(\vec\nabla\psi\times\vec\nabla\zeta\cdot\vec\nabla\theta)^{-1}$. The following identities,
\[
    \frac{\vec\nabla\psi\times\vec B}{B^2}=R^2\vec\nabla\zeta-I(\psi)\frac{\vec{B}}{B^2}
\]
and
\[
    \Bigl\langle\vec\nabla\cdot\vec A\Bigr\rangle=\frac{1}{V'}\frac{\partial}{\partial\psi}\Bigl(V'\langle\vec A\cdot\vec\nabla\psi\rangle\Bigr)
\]
are used, where $V$ is the volume element of the flux tube and $V'\equiv \partial V/\partial\psi$.

Finally, by eliminating the expression $\langle\delta q(\vec{u}_E\cdot\vec{\nabla})\psi\rangle$ from Eqs.~(\ref{eqn8}) and (\ref{eqn9}), we obtain the desired momentum theorem:
\begin{equation}\label{eqn11}
    \frac{\partial}{\partial t}\left(N\left\langle\frac{1}{B^2}\vec{\nabla}\Phi\cdot\vec{\nabla}\psi\right\rangle+\frac{\langle\delta q^2\rangle}{2B^2\frac{\partial}{\partial \psi}\left(\frac{N}{B^2}\right)}\right)=-\frac{1}{2}\left[\frac{\partial}{\partial\psi}\left(\frac{N}{B^2}\right)\right]^{-1}\frac{1}{V'}\frac{\partial}{\partial \psi}\Bigl(V'B^2\langle\delta q_M^2\vec{u}_E\cdot\vec{\nabla}\psi\rangle\Bigr),
\end{equation}
where we have assumed $N$ is a flux function (of $\psi$ only).

The LHS consists of time evolution of zonal flow momentum (the first term) and the negative of pseudo-momentum (the second term). Here, when $\Phi$ is a flux-function, i.e., a function of $\psi$ only, we can show that
\begin{equation}\label{eqn12}
    \frac{\vec{\nabla}\phi\cdot\vec{\nabla}\psi}{B^2}=\frac{RB_\theta}{cB_\zeta}u_\theta
\end{equation}
and
\begin{equation}\label{eqn13}
    \left\langle\frac{\vec{\nabla}\phi\cdot\vec{\nabla}\psi}{B^2}\right\rangle=-\left\langle\frac{(RB_\theta)^2}{B}\right\rangle\left(\frac{E_r}{RB_\theta}\right).
\end{equation}
So this expression is proportional to the poloidal component of $E\times B$ flow and closely related to the $E\times B$ flow shearing rate expression in toroidal geometry. \cite{HahmPoP95} The RHS represents the radial flux of the magnetic potential enstrophy density which can be regarded as turbulence spreading.

Therefore, Eq.~(\ref{eqn11}) describes a conservation of the total momentum which is a sum of the zonal flow momentum and the wave momentum in the absence of turbulence spreading, forcing and dissipation. It also implies that stationary turbulence (for which the second term of the LHS vanishes) cannot accelerate the zonal flow when turbulence spreading, forcing and dissipations are negligible. This is the reason behind another name ``non-acceleration theorem" for the Charney-Drazin momentum theorem. \cite{CharneyJGR61} Most literature to date has emphasized the balance between two terms on the LHS. \cite{DiamondPPCF08,KosugaNF13,GurcanJPA15}

In this work, we found that after a local saturation of turbulence, turbulence spreading radially carries zonal flows alongside. The momentum theorem (Eq.~(\ref{eqn11})) states that turbulence spreading (the RHS) can induce the generation of the perpendicular momentum which consist of the zonal flow (the 1\textsuperscript{st} term on the LHS) and the wave pseudo-momentum (the 2\textsuperscript{nd} term on the LHS) which is also known as the wave action density.

\section{Gyrokinetic Simulations}\label{sec:simulation}

In this section, simulation results using \texttt{gKPSP} \cite{KwonNF12,KwonCPC17,SeoCPC21} code are presented. \texttt{gKPSP} solves the 5-D $\delta f$ nonlinear gyrokinetic governing equations using particle-in-cell approach. Global nonlinear gyrokinetic simulations are performed to investigate the interplay between turbulence spreading and zonal flow. 

Role of PV dynamics which governs both zonal flow and turbulence spreading has been mostly elucidated in the cold ion limit $(T_i/T_e\ll1)$ where the formulation simplifies considerably. These include a work showing that drift waves can convert parallel compression into perpendicular zonal flows via a coupling to ion sound wave. \cite{WangPPCF12} Notable exceptions are Ref.~\onlinecite{McDevittPoP10} and Ref.~\onlinecite{YiNF19}. Ref.~\onlinecite{McDevittPoP10} has provided an expression of the PV which include finite Larmor radius (FLR) effects which occur for $k_\perp\rho_i\sim1$. As expected, the expression is very complex and not really appropriate for physics validation purposes. Ref.~\onlinecite{YiNF19} has identified an important role of parallel compressibility in PV and zonal flow dynamics from \texttt{gKPSP} simulations for $T_i=T_e$. However, role of turbulence spreading has not been addressed. A previous \texttt{gKPSP} simulation for KSTAR L-mode like parameters with $T_i/T_e\simeq 1$ reported that turbulence accompanies zonal flows as it spreads radially. \cite{YiPoP24} However, comparisons to theory have not been attempted.

With this in mind, we performed global linear gyrokinetic simulations at $T_i/T_e=0.1$ and compared it with $T_i/T_e=1.0$ case. $R/L_{T_i}$ values were adjusted so that the linear growth rates have similar magnitudes. The growth rate and real frequency spectra are shown in Fig.~\ref{fig1}. It is apparent that, as we reduce the ion temperature, the growth rate spectrum shifts to lower $k_\perp\rho_i$ range, satisfying the $k_\perp\rho_i\ll1$ assumption in the cold ion limit. The equilibrium profiles of the nonlinear simulation are shown in Fig.~\ref{fig2}. The functional forms of the profiles are given by the normalized gradient scale lengths
\begin{equation}
    \frac{R}{L_A}=\kappa_A\Biggl[1-\cosh^{-2}\left(\frac{r-(r_0-\Delta r)}{a\Delta A}\right)-\cosh^{-2}\left(\frac{r-(r_0+\Delta r)}{a\Delta A}\right)\Biggr]
\end{equation}
in the region $\left[r_0-\Delta r\le r\le r_0+\Delta r\right]$ and zero in the rest of the domain. Here, $A$ represents either density $n$ or temperature $T$. This choice has been made to clearly demonstrate turbulence spreading into the linearly stable zones. The values for each parameter are $\kappa_T=2.70$, $\kappa_n=2.23$, $r_0=0.5a$, $\Delta r=0.25a$, and $\Delta T=\Delta n=0.02$. Turbulence spreading (TS) has been observed in simulations for reversed magnetic shear plasmas in the inner core, \cite{YagiPPCF06, DengPoP09} and for the edge-scrape-off-layer (SOL) coupling regions. \cite{LiNF23} Experimental evidences are also numerous. \cite{HahmJKPS18,GrenfellNF20,EstradaNF11,KobayashiPRL22,KhabanovNF24} So we expect TS will persist even for more realistic profiles and geometry, although a quantitative assessment of such effects is beyond the scope of this paper. The gradients in the linearly unstable region are well beyond the Dimits shift \cite{DimitsPoP00} (effective nonlinear threshold for ion temperature gradient instability) regime. Electrons are assumed to be adiabatic, and ion temperature is set to be ten times lower than that of electron $(T_i/T_e=0.1)$. The rest of the plasma parameters are largely based on the CYCLONE base case: \cite{DimitsPoP00} a concentric circular-shaped deuterium plasma with $R_0=167\ cm$, $a=60.12\ cm$, $B_0=2.0\ T$. The radial safety factor ($q$) profile is given as $q=0.86-0.44\epsilon+19.4\epsilon^2$ where $\epsilon\equiv r/R_0$, which yields $q=1.41$ and $\hat{s}=0.84$ at $r/a=0.5$. The central density and temperatures are $n_{i0}=n_{e0}=5.0\times10^{13}\ cm^{-3}$ and $T_{e0}=10\ T_{i0}=4.2\ keV$.

\begin{figure*}
\includegraphics[width=\textwidth]{linear_scan.png}
\caption{\label{fig1}Linear gyrokinetic simulation results. (a) shows the linear growth rate spectra of both the cold ion case $(T_i/T_e=0.1)$ and the reference case ($T_i/T_e=1.0)$, and (b) shows the real frequency spectra. Blue lines correspond to the cold ion results, and black lines correspond to the reference results. The normalized ion temperature gradient scale lengths were adjusted so that the two cases exhibit similar linear growth rates.}
\end{figure*}

\begin{figure*}
\includegraphics[width=\textwidth]{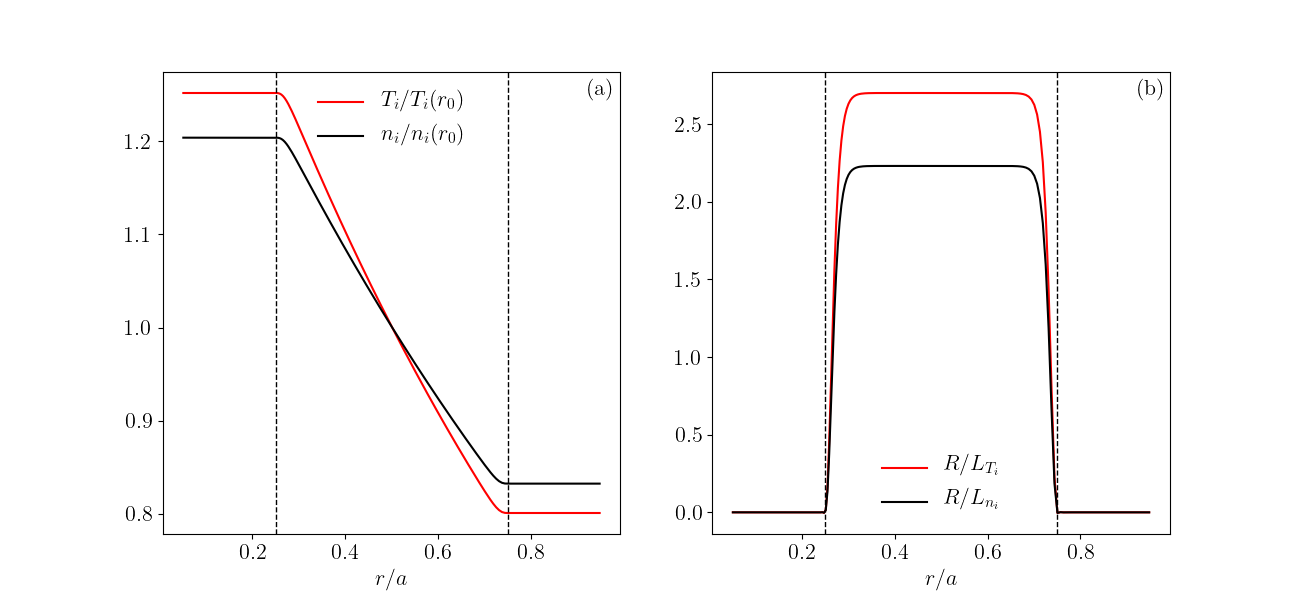}
\caption{\label{fig2}Profiles used for \texttt{gKPSP} simulation. (a) shows the ion temperature and density profiles normalized to the value at the core, and (b) shows the normalized gradient scale lengths. The normalized gradient scale lengths are zero in the region where $r/a<0.24$ and $r/a>0.76$ indicated by vertical dashed lines.}
\end{figure*}

\begin{figure*}
\includegraphics[trim={2cm 0 3cm 0},clip,width=1.0\textwidth,]{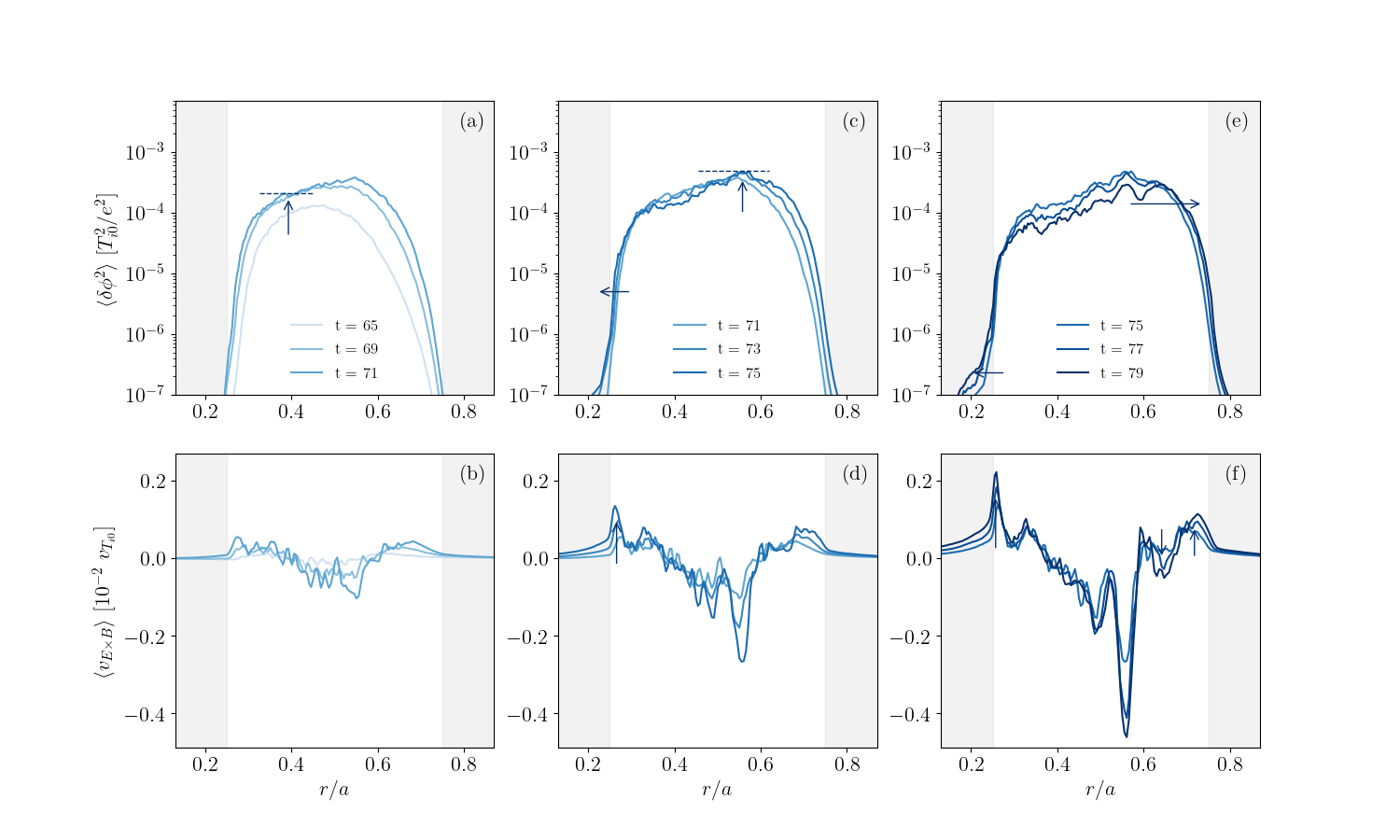}
\caption{\label{fig3}Evolution of electrostatic potential fluctuation intensity profile and zonal flow profile. (a), (c) and (e) show the time trace of flux-surface-averaged $\delta\phi^2$ profiles, whereas (b), (d) and (f) show the corresponding flux-surface-averaged zonal flow profiles. The time point values normalized by $R_0/v_{T_{i0}}$ are indicated by the legends. The gray shaded areas correspond to regions where the normalized gradient scale length is zero.}
\end{figure*}

The evolutions of the turbulence intensity and zonal flow profiles, shown in Fig.~\ref{fig3}, occur in three distinct phases. In the first phase, which corresponds to Fig.~\ref{fig3}(a) and (b), nonlinear saturation occurs in the strongly unstable region. In Fig.~\ref{fig3}(a), this is observed at $r/a=0.4$ around $t=71\ [R_0/v_{T_{i0}}]$. The saturated intensity level is marked with a horizontal dashed line. At the same time, turbulence continues to grow in other radial positions, as shown by the difference between the intensity profiles at $t=69-71\ [R_0/v_{T_{i0}}]$.

The second phase, shown in Fig.~\ref{fig3}(c) and (d), is characterized by the turbulence reaching global saturation during $t=73-75\ [R_0/v_{T_{i0}}]$ (heat flux level peaks at $t=73.5\ [R_0/v_{T_{i0}}]$). The saturated level is indicated with a dashed line in Fig.~\ref{fig3}(c). Concurrently, turbulence spreads from the earlier nonlinear saturation region ($r/a=0.4$) into the region with zero turbulence drive ($r/a<0.25$). The zonal flow profile also exhibits correlated evolution in that region (see Fig.~\ref{fig3}(d)). Although turbulence saturation at $r/a\sim0.3$ occured at $t=71\ [R_0/v_{T_{i0}}]$, continued zonal flow generation is observed well past that time. The radial position and extent of the zonal flow profile evolution show similarity with that of the spreading turbulence.

The third phase involves turbulence spreading on a global scale (see Fig.~\ref{fig3}(e) and (f)). We note that after the nonlinear saturation of turbulence in the strongly driven region, its intensity gradually settle down to a slightly lower level as observed in previous nonlinear gyrokinetic simulations. \cite{LinPRL02,HahmPPCF04} Continuing turbulence spreading and correlated zonal flow generation around $r/a=0.3$ region are observed. Additionally, turbulence originating from the region of the second phase's nonlinear saturation spreads into neighboring regions. From the evolution of the turbulence intensity profile in Fig.~\ref{fig3}(e), turbulence spreading from $r/a=0.55$ to $r/a=0.75$ can be identified. At the same time, zonal flow generation in the corresponding region is also clearly observed in Fig.~\ref{fig3}(f). In conclusion, our simulation results indicate turbulence spreading carries zonal flow alongside to different radial locations, even into the linearly stable zones.

\begin{figure*}
\includegraphics[trim={0 0 0 0},clip,width=\textwidth]{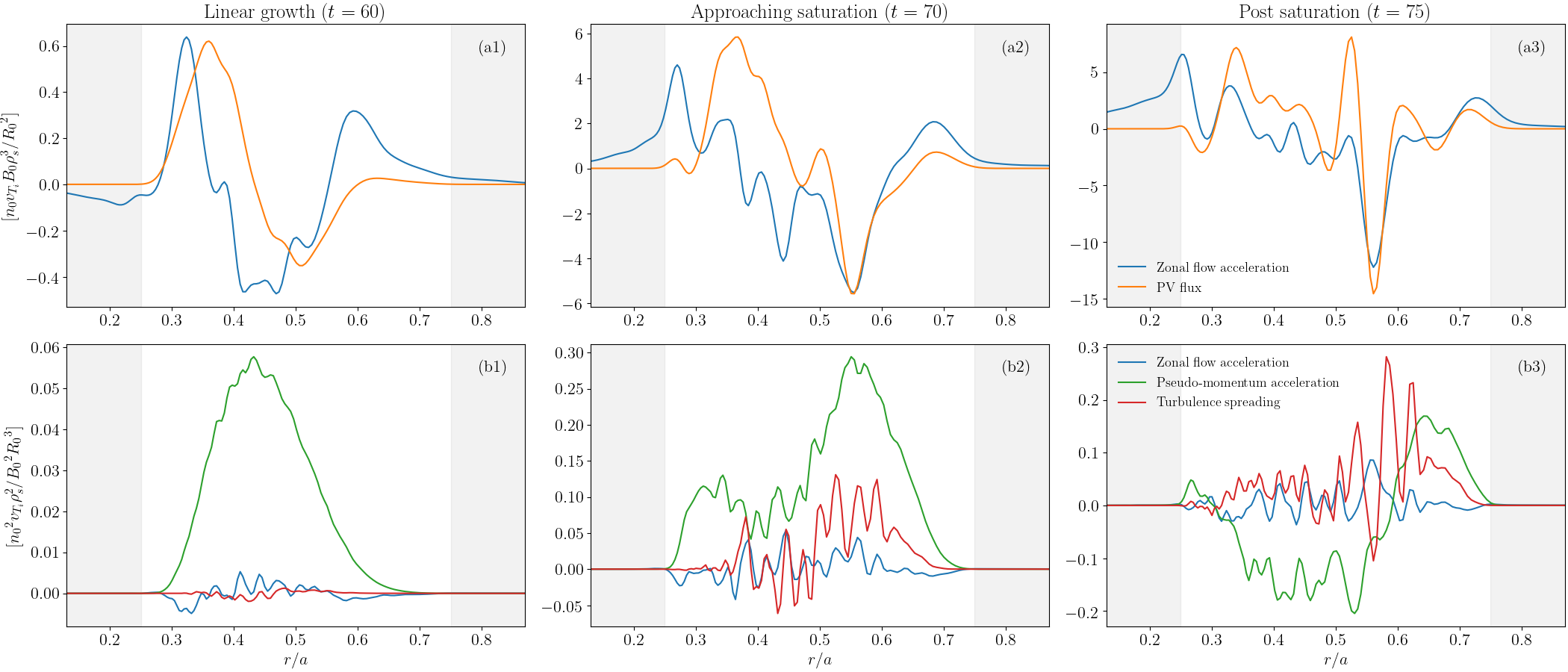}
\caption{\label{fig4}Comparative plots of each term of the momentum theorem equation at three different time point in \texttt{gKPSP} simulation. (a1)-(a3) compares the profiles of the zonal flow acceleration (blue) and the PV flux (orange), and (b1)-(b3) compares the zonal flow acceleration profile (blue), pseudo-momentum acceleration profile (green), and the turbulence spreading profile (red). Associated physics discussions are given in the latter part of Sec.~III. The first column corresponds to the linear growth phase of the simulation, the second column corresponds to a time point close to the nonlinear saturation, and the third column is after the nonlinear saturation.}
\end{figure*}

Our \texttt{gKPSP} simulation results are analyzed and discussed in the context of predictions from the momentum theorem in toroidal geometry. \cite{HahmPoP24} The profiles of the relevant quantities at three distinct phases are plotted in Fig.~\ref{fig4}. Fig.~\ref{fig4}(a1)-(a3) compare the zonal flow acceleration term (the LHS of Eq.~\ref{eqn8}) and the radial potential vorticity flux term (the RHS of Eq.~\ref{eqn8}). Two quantities are semi-quantitatively similar in both magnitudes and profiles, indicating a high relevance of PV flux in zonal flow acceleration. Fig.~\ref{fig4}(b1)-(b3) compares each of the terms of the momentum theorem. Eq.~(\ref{eqn11}) is rearranged in the following form because $\partial(N/B^2)/\partial\psi$ is small and negative:
\begin{eqnarray}\label{eqn15}
    \frac{\partial}{\partial t}\left(N\left\langle\frac{1}{B^2}\vec{\nabla}\Phi\cdot\vec{\nabla}\psi\right\rangle\right)\frac{\partial}{\partial\psi}\left(\frac{N}{B^2}\right)+\frac{\partial}{\partial t}\left(\frac{\langle\delta q^2\rangle}{2B^2}\right)=-\frac{1}{2}\frac{1}{V'}\frac{\partial}{\partial \psi}\Bigl(V'B^2\langle\delta q_M^2\vec{u}_E\cdot\vec{\nabla}\psi\rangle\Bigr).
\end{eqnarray}
The first term on the LHS, which is the zonal flow acceleration term, is represented by the blue curve, and the second term on the LHS, which is the drift wave pseudo-momentum term, is plotted as the green curve. The turbulence spreading term on the RHS is drawn in red.

In the linear growth phase (Fig.~\ref{fig4}(b1)), the drift wave pseudo-momentum dominates as it is proportional to the turbulence intensity while zonal flow generation and turbulence spreading are in their initial stage.

As we approach nonlinear saturation (Fig.~\ref{fig4}(b2)), similarity between the zonal flow acceleration profile and turbulence spreading profile in the radial region $r/a<0.5$ is observed. In the $r/a>0.5$ region where turbulence is still growing, turbulence spreading term starts to increase as well. We note that the higher magnitude of drift wave pseudo-momentum term compared to other two terms in Eq.~(\ref{eqn15}) is probably due to other effects not included in the derivation thereof. \cite{HahmPoP24} In particular, FLR terms have been ignored due to the cold ion approximation.

After nonlinear saturation shown in Fig.~\ref{fig4}(b3), the role of turbulence spreading becomes more conspicuous. As turbulence intensity adjusts to a lower value, the drift wave pseudo-momentum profile takes negative values. The region where turbulence spreading profile resembles zonal flow profile broadens ($r/a<0.55$). This is a clear indication that turbulence spreading and the zonal flow generation occur simultaneously as predicted by the momentum theorem, and this behavior continues well after nonlinear saturation, as shown in the third column of Fig.~\ref{fig4}. The most important conclusion of this paper is an understanding of the relation between zonal flow generation and turbulence spreading in the context of the extended momentum theorem as discussed in Sec.~\ref{sec:background}.

\section{Discussion and Conclusion}\label{sec:conclusion}

In conclusion, our nonlinear gyrokinetic simulations show that after a local saturation of turbulence, turbulence spreading transports zonal flows radially, extending into the linearly stable regions. A theoretical understanding of the results has been gained in the context of an extended momentum theorem. \cite{HahmPoP24}

We note that the momentum theorem has also provided an insight on the problem of $E\times B$ vortex generation inside a large magnetic island previously. \cite{YoonNF24} Since the equilibrium profiles inside a magnetic island get usually flattened considerably, one does not expect any linear microinstabilities there. Nevertheless, turbulence and $E\times B$ vortex have been observed experimentally. \cite{EstradaNF16} Ref.~\onlinecite{YoonNF24} has derived the momentum theorem in a magnetic island geometry and has shown that there can be a dominant balance between turbulence spreading through X-points of an island and $E\times B$ vortex flow generation inside. This is consistent with the experimental observation \cite{IdaPRL18,ChoiNatComm21} and provides an explanation how turbulence and $E\times B$ vortex flow exist inside an island even in the absence of any linear instability there. 

Possible future studies include an extension of our studies to higher values of $T_i/T_e$. A previous gyrokinetic simulation for $T_i/T_e\simeq 1$ has already reported a trend which is similar to our results. \cite{YiPoP24} For comparison to theory, elucidating the role of a term proportional to $\vec{\nabla} B\times\vec{\nabla}\delta P_i$ which was shown to modify the PV evolution in Ref.~\onlinecite{McDevittPoP10} will be worth an effort. This term is analogous to the baroclinic vector known to modify the Charney-Drazin theorem in GFD. \cite{PedloskyBook} Another obvious extension is to consider an electromagnetic turbulence in simulations and theory. On this issue, there exists global gyrokinetic simulation which exhibits radial convection of entropy associated with electromagnetic ITG turbulence and zonal modes. \cite{MasuiNF22}

\begin{acknowledgments}
The authors would like to acknowledge fruitful discussions with A. Ishizawa, G. J. Choi, Lu Wang, and Weixin Guo. This work was supported by the National Research Foundation of Korea (NRF) grant funded by the Korea Government (MSIT) (Grant No. 2023R1A2C100773512), and by the R\&D Program through the Korea Institute of Fusion Energy (KFE) funded by the Ministry of Science and ICT of the Republic of Korea (EN2541). The simulations presented in this work were performed using the HPC resources from KFE KAIROS. The authors also gratefully acknowledge The Research Institute of Energy and Resources and The Institute of Engineering Research at Seoul National University.
\end{acknowledgments}

\bibliography{reference}

\clearpage

\end{document}